\documentstyle[11pt]{article}

\makeatletter
\@addtoreset{equation}{section}
\makeatother

\makeatletter
\def\citen#1{%
\edef\@tempa{\@ignspaftercomma,#1, \@end, }% ignore spaces in parameter list
\edef\@tempa{\expandafter\@ignendcommas\@tempa\@end}%
\if@filesw \immediate \write \@auxout {\string \citation {\@tempa}}\fi
\@tempcntb\m@ne \let\@h@ld\relax \def\@citea{}%
\@for \@citeb:=\@tempa\do {\@cmpresscites}%
\@h@ld}
\def\@ignspaftercomma#1, {\ifx\@end#1\@empty\else
   #1,\expandafter\@ignspaftercomma\fi}
\def\@ignendcommas,#1,\@end{#1}
\def\@cmpresscites{%
 \expandafter\let \expandafter\@B@citeB \csname b@\@citeb \endcsname
 \ifx\@B@citeB\relax % undefined
    \@h@ld\@citea\@tempcntb\m@ne{\bf ?}%
    \@warning {Citation `\@citeb ' on page \thepage \space undefined}%
 \else%  defined
    \@tempcnta\@tempcntb \advance\@tempcnta\@ne
    \setbox\z@\hbox\bgroup % check if citation is a number:
    \ifnum0<0\@B@citeB \relax
       \egroup \@tempcntb\@B@citeB \relax
       \else \egroup \@tempcntb\m@ne \fi
    \ifnum\@tempcnta=\@tempcntb % Number follows previous--hold on to it
       \ifx\@h@ld\relax % first pair of successives
          \edef \@h@ld{\@citea\@B@citeB }%
       \else % compressible list of successives
          \edef\@h@ld{\hbox{--}\penalty\@highpenalty
            \@B@citeB }%
       \fi
    \else   %  non-successor--dump what's held and do this one
       \@h@ld\@citea\@B@citeB
       \let\@h@ld\relax
 \fi\fi%
 \def\@citea{,\penalty\@highpenalty\hskip.13em plus.1em minus.1em}%
}
\def\@citex[#1]#2{\@cite{\citen{#2}}{#1}}%
\def\@cite#1#2{\leavevmode\unskip
  \ifnum\lastpenalty=\z@\penalty\@highpenalty\fi% highpenalty before
  \ [{\multiply\@highpenalty 3 #1%             % triple-highpenalties within
      \if@tempswa,\penalty\@highpenalty\ #2\fi % and before note.
    }]\spacefactor\@m}
\makeatother

\def\half{{\textstyle{1\over2}}}
\def\dalemb#1#2{{\vbox{\hrule height .#2pt
	\hbox{\vrule width.#2pt height#1pt \kern#1pt
			\vrule width.#2pt}
				\hrule height.#2pt}}}

\def\apm{{\alpha'}}

\let\a=\alpha \let\b=\beta

\def\nn{\nonumber} \def\bd{\begin{document}} \def\ed{\end{document}}
\def\ds{\documentstyle} \let\fr=\frac \let\bl=\bigl \let\br=\bigr
\let\Br=\Bigr \let\Bl=\Bigl
\let\bm=\bibitem
\let\na=\nabla
\let\pa=\partial \let\ov=\overline
\def\ft#1#2{{\textstyle{{\scriptstyle #1}\over {\scriptstyle #2}}}}
\def\fft#1#2{{#1 \over #2}}
\def\del{\partial}
\newcommand{\be}{\begin{equation}}
\newcommand{\ee}{\end{equation}}
\def\ba{\begin{array}}
\def\ea{\end{array}}
\def\sst#1{{\scriptscriptstyle #1}}
\newcommand{\ho}[1]{$\, ^{#1}$}
\newcommand{\hoch}[1]{$\, ^{#1}$}
\newcommand{\bea}{\begin{eqnarray}}
\newcommand{\eea}{\end{eqnarray}}
\newcommand{\ra}{\rightarrow}
\newcommand{\lra}{\longrightarrow}
\newcommand{\Lra}{\Leftrightarrow}
\newcommand{\ap}{\alpha^\prime}
\newcommand{\bp}{\tilde \beta^\prime}
\newcommand{\tr}{\,{\rm tr}\,}
\newcommand{\Tr}{{\rm Tr} }
\newcommand{\NP}{Nucl. Phys. }
\newcommand{\tamphys}{\it
Center for Theoretical Physics, Department of Physics\\
Texas A\&M University, College Station, Texas 77843--4242}
\newcommand{\ens}{\it Laboratoire de Physique Th\'eorique de l'\'Ecole
Normale Sup\'erieure\hoch{4}\\
24 Rue Lhomond - 75231 Paris CEDEX 05, France}

\begin{document}

\rightline{CTP-TAMU-46/97}
\rightline{RU97-7-B}
\rightline{LPTENS-97/52}
\rightline{hep-th/9711089}
\rightline{November 1997}

\vspace{10pt}

\begin{center}
{ \large {\bf Gauge Dyonic Strings and Their Global Limit}}

\vspace{24pt}

M.J.~Duff\hoch{\star\,1}
James T.~Liu\hoch{\dagger\,2}
H.~L\"u,\hoch{\ddagger}
and C.~N.~Pope\hoch{\star\,3}

\vspace{10pt}

{\hoch{\star}\tamphys}

\bigskip

\hoch{\dagger}{\it Department of Physics, The Rockefeller University\\
1230 York Avenue, New York, NY 10021-6399}

\bigskip

\hoch{\ddagger}\ens

\vspace{20pt}

\underline{ABSTRACT}

\end{center}

We show that six-dimensional supergravity coupled to tensor and
Yang-Mills multiplets admits not one but two different theories as
global limits, one of which was previously thought not to arise as a
global limit and the other of which is new. The new theory has the
virtue that it admits a global anti-self-dual string solution obtained
as the limit of the curved-space gauge dyonic string, and
can, in particular, describe tensionless strings. We speculate that this
global model can also represent the worldvolume theory of coincident
branes. We also discuss the Bogomol'nyi
bounds of the gauge dyonic string and show that, contrary to
expectations, zero eigenvalues of the Bogomol'nyi matrix do not lead to
enhanced supersymmetry and that negative tension does not necessarily
imply a naked singularity.

{\vfill\leftline{}\vfill
\vskip  10pt
\footnoterule
{\footnotesize \hoch{1} Research supported in part by NSF Grant
PHY-9411543 \vskip -12pt} \vskip 14pt
{\footnotesize \hoch{2} Research supported in part by the U.S. DOE
under grant no. DOE-91ER40651-TASKB \vskip -12pt} \vskip 14pt
{\footnotesize \hoch{3} Research supported in part by the U.S. DOE
under grant no. DE-FG03-95ER40917 \vskip -12pt} \vskip 14pt
{\footnotesize \hoch{4} Unit\'e Propre du Centre National de la Recherche
Scientifique, associ\'ee \`a l'\'Ecole Normale Sup\'erieure \vskip -12pt}
                       \vskip 10pt
{\footnotesize \hoch{\phantom{4}} et \`a l'Universit\'e de
Paris-Sud \vskip -12pt}}

\pagebreak
\setcounter{page}{1}

%%%%%%%%%%%%%%%%%%%%%%%%%%%%%%%%%%%%%%%%%%%%%%%%%%%%%%%%%%%%%%%%%%%%%%%%%
\section{Introduction}

This paper is devoted to certain properties of the six-dimensional
gauge dyonic string \cite{gaugestring} and in particular to its
global limit in which it becomes anti-self dual. An important special
case corresponds to the tensionless string,
which has been the subject of much interest lately
\cite{Witten,Duffelectric,Lupope,Strominger,Pt,md,mg,Becker,Witten2,%
Seibergwitten,asy,Ganor,gaugestring}, especially in the
context of phase transitions
\cite{DMW,Seibergwitten,gaugestring,Ibanez}.%
\footnote{But note that, contrary
to some claims in the literature, the tensionless string corresponds
to the (quasi)-anti-self-dual limit of the dyonic string of
\cite{Ferrara}, where the string couples dominantly to the 3-form
field strength of the tensor matter multiplet, and not the self-dual
string of \cite{Dufflu} where the string couples only to the 3-form
field strength of the gravity multiplet.}
This global limit is
particularly interesting because one might then expect to be able to
find an anti-self-dual string solution by directly solving the global
supersymmetric theory in six-dimensions
\cite{Seibergwitten} describing an anti-self-dual tensor multiplet
coupled to Yang-Mills. However, an apparently paradoxical claim was
made in
\cite{Bergshoeff:1996qm} that no such global limit exists.  Here we
resolve the paradox, and show that not only does the limit exist but
that there are in fact two different limits, each giving different
globally supersymmetric theories.  One of these is the theory
constructed in
\cite{Bergshoeff:1996qm}, which we shall refer to as the ``BSS theory''.
The other flat-space theory, which for reasons described below we
shall refer to as the ``interacting theory'', appears to be new, and
admits an anti-self-dual string solution which can indeed be obtained as
the flat-space limit of the dyonic string of the supergravity
theory.

    A surprising feature of the BSS theory constructed in
\cite{Bergshoeff:1996qm} is that there is an asymmetry in the
interactions between the Yang-Mills multiplet and the anti-self-dual
tensor multiplet.  In particular, the Yang-Mills multiplet satisfies
free equations of motion, whereas the equations of motion for the
tensor multiplet do involve couplings to the Yang-Mills fields.  By
contrast, the interactions in the ``interacting" theory obtained in
the present paper here are more symmetrical, in that they occur in all
the equations of motion.  Interestingly, however, the additional
interaction terms of the new theory cancel in the special case of its
anti-self-dual string solution, and so the same configuration is also a
solution of the BSS theory. Curiously, however, it is not tensionless
in that theory, and indeed the BSS theory is inappropriate for
describing any tensionless string solution.

    Another intriguing aspect of the gauge dyonic string concerns the
counter-intuitive relations between its Bogomol'nyi bound, unbroken
supersymmetry and its singularity structure \cite{gaugestring}. We confirm:

(1) The dyonic string continues to preserve just half of the
supersymmetry even in the tensionless limit, notwithstanding the
standard Bogomol'nyi argument that a BPS state with vanishing central
charge leads to completely unbroken supersymmetry.

(2) A solution with {\it negative} tension can be completely
non-singular, contrary to the folk-wisdom that negative mass
necessarily implies naked singularities.

     Finally, six dimensional global models are also important as
fivebrane worldvolume theories
\cite{Dijkgraaf,Seiberg,Schwarz,Bergshoeff:1996qm} and as the 
worlvolume theories of coincident higher-dimensional branes with six 
dimensions in common \cite{Intriligator,Hanany}. We
speculate that the interacting
anti-self-dual-tensor Yang-Mills system is indeed such a worldvolume
theory.  Hence the global gauge anti-self-dual string,
and in particular the tensionless string, may also be regarded as a
string on the worldvolume.   In the case of the tensionless
string, in the limit as the size
$\rho$ of the Yang-Mills instanton shrinks to zero, one recovers the
global limit of the neutral tensionless string
\cite{Ferrara,gaugestring} which is also a solution of the $(2,0)$
theory that resides on the worldvolume of the
$M$-theory fivebrane. It is curious, therefore, that we find in this
limit that the tension really is zero, as opposed to the infinite
tension of the string solution of the free $(2,0)$ theory \cite{Perry,Howe}.

\section{$N=1$ supergravity and the gauge dyonic string}

     The low-energy $D=6$ $N=(1,0)$ supergravity is generated by a
pair of symplectic Majorana-Weyl spinors $\epsilon$ transforming in
the $2$ of $Sp(2)$.  This theory has the unusual feature in that the
antisymmetric tensor breaks up into self-dual and anti-self-dual
components.  The basic supergravity theory consists of the graviton
multiplet $(g_{\mu\nu},\psi_\mu,B_{\mu\nu}^+)$ coupled to $n_T$ tensor
multiplets $(B_{\mu\nu}^-,\chi,\phi)$.  When $n_T=1$, corresponding to
the heterotic string compactified on $K3$, these multiplets may be
combined, yielding a single ordinary antisymmetric tensor
$B_{\mu\nu}$.

    We are interested, however, in the general case with $n_T$ tensor
multiplets coupled to an arbitrary number of vector multiplets
$(A_\mu,\lambda)$.  Due to the presence of chiral antisymmetric tensor
fields, there is no manifestly covariant Lagrangian formulation of
this theory.  Nevertheless, the equations of motion may be
constructed, and were studied in \cite{RomansSD,Sagnotti}.  With $n_T$
tensor multiplets, there are $n_T$ scalars parametrizing the coset
$SO(1,n_T)/SO(n_T)$.  This may be described in terms of a $(n_T+1)
\times (n_T+1)$ vielbein transforming as vectors of both $SO(1,n_T)$
and $SO(n_T)$.  Following the conventions of \cite{Sagnotti}, the
vielbein may be decomposed as
\begin{equation}
V=\left[\matrix{V_+\cr V_-}\right]=
\left[\matrix{v_0&v_M\cr x^m{}_0&x^m{}_M}\right]\ ,
\end{equation}
satisfying the condition $V^{-1}=\eta V^T\eta$ where
$\eta$ is the $SO(1,n_T)$ metric, $\eta=\hbox{diag}(1,-I_{n_T})$.
Below, we use indices $r,s,\ldots=\{0,M\}$ to denote $SO(1,n_T)$ vector
indices.  The composite $SO(n_T)$ connection is then given by
\begin{eqnarray}
S_\mu^{[mn]}&=&(\partial_\mu V_-^{\vphantom{T}}\eta V_-^T)^{[mn]}\nonumber\\
&=&-x^m{}_0\partial_\mu x^n{}_0+x^m{}_M\partial_\mu x^n{}_M\ ,
\end{eqnarray}
so that the fully covariant derivative acting on $SO(n_T)$ vectors is given
by ${\cal D}_\mu=\nabla_\mu+S_\mu$.

To describe the combined supergravity plus tensor system, we introduce
$(n_T+1)$ antisymmetric tensors $B_{\mu\nu}$ transforming as a vector of
$SO(1,n_T)$.  In the presence of Yang-Mills fields, the three-form field
strengths pick up a Chern-Simons coupling
\begin{equation}
{\cal H}=dB+c\,\omega_3\ ,\label{csterms}
\end{equation}
where $\omega_3=AdA+{2\over3}A^3$,
so that $d{\cal H}=c\tr F^2$.  The constants $c$ form a $(n_T+1)\times n_V$
matrix where $n_V$ is the number of vector multiplets%
\footnote{For non-abelian gauge fields, instead of having $n_V$ independent
quantities, there is a single set of $c$'s for each factor of
the gauge group.}.
Note that this
coupling of the vector and tensor multiplets is dictated by supersymmetry
and encompasses both tree-level and one-loop Yang-Mills corrections.
Furthermore, the supersymmetry guarantees that there are no
higher-loop corrections.
The vielbein is then used to transform the field strengths ${\cal H}$ into
their chiral components $H=v_r{\cal H}^r$ and $K^m=x^m{}_r{\cal H}^r$
so that the (anti-)self-duality conditions for the tensors become
$H = *H$ and $K^m = -*\!K^m$.

With the above conventions, the bosonic equations of motion are
\begin{eqnarray}
G_{\mu\nu}\equiv
R_{\mu\nu}-{\textstyle{1\over2}}g_{\mu\nu}R&=&T_{\mu\nu}\nonumber\\
{\cal D}_\mu P^{m\,\mu}&=&-\ft{\sqrt{2}}{3}H_{\mu\nu\rho}K^{m\,\mu\nu\rho}
-\ft1{\sqrt{2}}x^m{}_rc^r\tr (F_{\mu\nu}F^{\mu\nu})\nonumber\\
dH &=& - \sst{\sqrt{2}}P^mK^m+v_rc^r\tr F^2\nonumber\\
(d\delta^{mn} + S^{mn})K^n &=& - \sst{\sqrt{2}}P^m H +
x^m{}_rc^r\tr F^2\nonumber\\
v_rc^rD^\mu F_{\mu\nu}&=&\sst{\sqrt{2}} P^{m\,\mu} x^m{}_rc^rF_{\mu\nu}
+H_{\nu\rho\sigma}v_rc^rF^{\rho\sigma}
+K^{m}_{\nu\rho\sigma}x^m{}_rc^rF^{\rho\sigma}\ ,
\label{eq:eom}
\end{eqnarray}
where
\begin{eqnarray}
P^m_\mu&=&\ft1{\sqrt{2}}(\partial_\mu V_+\eta V_-^T)^m\nonumber\\
&=&\ft1{\sqrt{2}}(x^m{}_0\partial_\mu v_0-x^m{}_M\partial_\mu v_M)\ ,
\end{eqnarray}
and $S$ and $P$ are the 1-forms, $S = S_\mu dx^\mu$, $P = P_\mu dx^\mu$.
The symmetric stress tensor is given by
\begin{equation}
T_{\mu\nu}=H_{\mu\rho\sigma}H_\nu{}^{\rho\sigma}
+K^{m}_{\mu\rho\sigma}K_\nu^{m\,\rho\sigma}
+2[P^m_\mu P^m_\nu-\half g_{\mu\nu}P_\rho^mP^{m\,\rho}]
+4v_rc^r\tr [F_{\mu\lambda}F_\nu{}^\lambda-{\textstyle{1\over4}}g_{\mu\nu}
F_{\lambda\sigma}F^{\lambda\sigma}]\ .
\label{eq:tmunu}
\end{equation}
For the antisymmetric tensors, Eqn.~(\ref{eq:eom}) along with the
(anti-)self-duality constraint may be viewed as the equivalent of the
combined Bianchi identities and equations of motion.  Finally, the fermionic
equations of motion are
\begin{eqnarray}
\gamma^{\mu\nu\rho}\nabla_\nu\psi_\rho &=&
-H^{\mu\nu\rho}\gamma_\nu\psi_\rho
+ \ft{i}{2}K^{m\,\mu\nu\rho} \gamma_{\nu\rho}\chi^m
-\ft{i}{\sqrt{2}}P_\nu^m\gamma^\nu\gamma^\mu\chi^m
-\ft1{\sqrt{2}}\gamma^{\sigma\tau}\gamma^\mu v_rc^r\tr F_{\sigma\tau}\lambda
\nonumber\\
\gamma^\mu\nabla_\mu\chi^m&=&
\ft{i}{2}K^{m\,\mu\nu\rho}\gamma_{\mu\nu}\psi_\rho
+\ft1{12} H_{\mu\nu\rho}\gamma^{\mu\nu\rho}\chi^m
+\ft{i}{\sqrt{2}}P^m_\nu\gamma^\mu\gamma^\nu\psi_\mu
-\ft{i}{\sqrt{2}}\gamma^{\mu\nu}x^m{}_rc^r\tr F_{\mu\nu}\lambda\nonumber\\
v_rc^r\gamma^\mu D_\mu\lambda &=&\ft1{\sqrt{2}}P_\mu^m\gamma^\mu
x^m{}_rc^r\lambda
-\ft1{2\sqrt{2}}v_rc^rF_{\lambda\tau}\gamma^\mu\gamma^{\lambda\tau}
\psi_\mu
-\ft{i}{2\sqrt{2}} x^m{}_rc^rF_{\mu\nu} \gamma^{\mu\nu}\chi^m
\nonumber\\
&&-\ft1{12}K^m_{\mu\nu\rho}x^m{}_rc^r\gamma^{\mu\nu\rho}\lambda\ .
\end{eqnarray}

In order to examine the Bogomol'nyi bound, we need the supersymmetry
variations for the fermionic fields:
\begin{eqnarray}
\delta\psi_\mu&=&[\nabla_\mu+{\textstyle{1\over4}}H_{\mu\nu\rho}
\gamma^{\nu\rho}]\epsilon\nonumber\\
\delta\chi^m&=&i[{\textstyle{1\over\sqrt{2}}}\gamma^\mu P^m_\mu
+{\textstyle{1\over12}}K_{\mu\nu\rho}^m\gamma^{\mu\nu\rho}]\epsilon\nonumber\\
\delta\lambda&=&-{\textstyle{1\over2\sqrt{2}}}F_{\mu\nu}\gamma^{\mu\nu}
\epsilon
\label{eq:susyF}
\end{eqnarray}
(given to lowest order).  For completeness, the bosonic fields transform
according to
\begin{eqnarray}
\delta e_\mu{}^a&=&-i\overline{\epsilon}\gamma^a\psi_\mu\nonumber\\
\delta B^r_{\mu\nu}&=&\eta^{rs}\overline{\epsilon}
[iv_s\gamma_{[\mu}\psi_{\nu]}
-{\textstyle{1\over2}}x^m{}_s \gamma_{\mu\nu} \chi^m]
+2c^r\tr A_{[\mu}\delta A_{\nu]}\nonumber\\
\delta v_r&=&x^m{}_r\overline{\epsilon}\chi^m\nonumber\\
\delta A_\mu&=&-{\textstyle{i\over\sqrt{2}}}\overline{\epsilon}\gamma_\mu
\lambda\ .
\label{eq:susyB}
\end{eqnarray}
Careful examination of Eqns.~(\ref{eq:susyF}) and (\ref{eq:susyB}) reveals
the intricate interplay between terms of various chiralities necessary to
maintain $D=6$ $N=(1,0)$ supersymmetry.  In particular,
$\epsilon$ is a chiral spinor satisfying $P_+\epsilon=0$ where
$P_\pm={1\over2}(1\pm\gamma^7)$ is the chirality projection in six
dimensions.  As a consequence, $H$ and $K$ satisfy the identities
\begin{eqnarray}
(H_{\mu\nu\lambda}\gamma^{\mu\nu\lambda})\epsilon&=&0\nonumber\\
(K^m_{\mu\nu\lambda}\gamma^{\mu\nu\lambda}\gamma_\alpha)\epsilon&=&0\ ,
\end{eqnarray}
which prove to be useful in manipulating Nester's form below.

\subsection{The gauge dyonic string solution}

It was shown in \cite{gaugestring} that the equations of motion
(\ref{eq:eom}) admit a gauge dyonic string solution carrying both self-dual
and anti-self-dual tensor charges.  Under an appropriate $SO(n_T)$ rotation,
the latter charge can be put in a single tensor component, so that we may
focus on the theory with $n_T=1$.  In this case,
corresponding to a compactified heterotic string, the self-dual and
anti-self-dual three-forms in the graviton and tensor multiplets
respectively may be combined together according to
\begin{equation}
H=\half e^{-\phi} (*{\cal H}+ {\cal H})\ ,\qquad
K=\half e^{-\phi} (*{\cal H}- {\cal H})\ ,
\end{equation}
where we have chosen a vielbein
\begin{equation}
V=\left[\matrix{\cosh\phi&\sinh\phi\cr \sinh\phi&\cosh\phi}\right]\ .
\end{equation}
For a simple gauge group, we pick the coupling vector $c$ to be
\begin{equation}
c={\alpha'\over 16}\left[\matrix{v+\tilde v\cr -v+\tilde v}\right]\ ,
\end{equation}
so that the ${\cal H}$ Bianchi identity and equation of motion, given in
Eqn.~(\ref{eq:eom}), may be rewritten as
\begin{equation}
d{\cal H}=\ft18 \alpha'\, v \tr F\wedge F\qquad
d(e^{-2\phi}*{\cal H})= \ft18 \alpha'\, \tilde v\tr F\wedge F\ .
\label{eq:hbiaeom}
\end{equation}

The gauge dyonic string is built around a single self-dual $SU(2)$ Yang-Mills
instanton in transverse space, and is given in terms of three parameters,
which are the electric and magnetic charges $Q$ and $P$ carried by the
string, and $\rho$ which is the scale parameter of the instanton.
Splitting the six-dimensional space into longitudinal $\mu,\nu=0,1$ and
transverse $m,n,\ldots=2,3,4,5$ components, the
gauge dyonic string solution is given by \cite{gaugestring}
\begin{eqnarray}
&&ds^2= e^{2A}\eta_{\mu\nu}dx^\mu dx^\nu+e^{-2A}dy^mdy^m\nonumber\\
&&{\cal H}_{mnp}=\half\epsilon_{mnpq}\partial_q H_1\qquad
{\cal H}_{\mu\nu m}=\half\epsilon_{\mu\nu}\partial_m H_2^{-1}\nonumber\\
&&e^{-\phi}=\sqrt{H_2/H_1}\qquad e^{-2A}=\sqrt{H_1H_2}\ ,
\label{eq:gdssol}
\end{eqnarray}
where $\epsilon_{01}=1$, $\epsilon_{2345}=1$.  The functions $H_1$ and $H_2$
are
\begin{equation}
H_1 = e^{\phi_0} + \fft{P(2\rho^2 +r^2)}{(\rho^2+r^2)^2}\ , \qquad
H_2 = e^{-\phi_0} + \fft{Q(2\rho^2 +r^2)}{(\rho^2+r^2)^2}\ ,
\label{aph}
\end{equation}
and are determined by the effect of the instanton source
\begin{equation}
F^a={2\rho^2\over(\rho^2+r^2)^2}\eta_{mn}^ady^m\wedge dy^n\ ,
\end{equation}
on the three-form tensor according to (\ref{eq:hbiaeom}).  (Note that
$\tr(F^2)= 2 F^a_{mn}\, F^{a mn}$.)
In particular, the
charges are thus given by $Q=2\apm \tilde v$ and $P=2\apm v$.
The mass per unit length of the dyonic string is given by
\be
2\pi \a'^2\, m= P e^{-\phi_0} + Q e^{\phi_0}\ .\label{massform}
\ee
This expression for the mass, and its relation to the Bogomol'nyi bound,
will be examined in detail in the following section.

    In the $\rho\to0$ limit, we recover the neutral dyonic string
obtained in \cite{Ferrara}.

\section{The Bogomol'nyi bound in six dimensions}

It is well known that the six-dimensional $N=(1,0)$ supersymmetry algebra
admits a single real string-like central charge, putting a lower bound on
the tension of the six-dimensional string.  Thus the tensionless string
only arises in the limit of vanishing central charge.  Before focusing on
the tensionless string, we examine the Bogomol'nyi mass bound in general and
determine the conditions for which it is satisfied.

For a string-like field configuration in six dimensions, we may
construct the supercharge per unit length of the string from the behavior
of the gravitino at infinity \cite{Dabholkar}
\begin{equation}
Q_\epsilon=\int_{\partial {\cal M}}\overline{\epsilon}
\gamma^{\mu\nu\lambda}\psi_\lambda d\Sigma^{\mu\nu}\ ,
\label{eq:supercharge}
\end{equation}
where ${\cal M}$ is the four-dimensional space transverse to the string.
We note that in writing the supercharge in terms of the gravitino, this
expression holds only up to the equations of motion.  It is for this
reason that, unlike in the global case, saturation of the Bogomol'nyi
bound alone is insufficient to guarantee that the bosonic background
solves the supergravity equations of motion.

Using Nester's procedure \cite{Nester,Dabholkar,Harveyliu}, we may take the
anticommutator of two supercharges to get
\begin{equation}
\{Q_\epsilon,Q_{\epsilon'}\}=\delta_\epsilon Q_{\epsilon'}
=\int_{\partial{\cal M}} N^{\mu\nu} d\Sigma_{\mu\nu}\ ,
\label{eq:qqanti}
\end{equation}
where
\begin{equation}
N^{\mu\nu}=\overline{\epsilon'}\gamma^{\mu\nu\lambda}
\delta_\epsilon\psi_\lambda=\overline{\epsilon'}\gamma^{\mu\nu\lambda}
[\nabla_\lambda+{\textstyle{1\over4}}H_{\lambda\rho\sigma}
\gamma^{\rho\sigma}]\epsilon
\end{equation}
is a generalized Nester's form.  Appealing to the supersymmetry algebra, we
then see that the mass and central charge per unit length of the
six-dimensional string is encoded in the surface integral of $N^{\mu\nu}$.
For a string in the $0$-$1$ direction, the ADM mass per unit length $M$
of the string is given by the asymptotic behavior of the metric
\begin{equation}
ds^2=(1-{GM\over2r^2}+\cdots)[-dt^2+dz^2]+(1+{GM\over2r^2}+\cdots)dy^idy^i\ ,
\end{equation}
where $r^2=y^iy^i$ is the transverse radial distance from the string.
Using this definition of the ADM mass, the surface integral of Nester's form
becomes
\begin{equation}
\int_{\partial{\cal M}}N^{\mu\nu}d\Sigma_{\mu\nu}
=2\pi^2 \, \epsilon^{\prime\dagger}[{GM\over2}-Z\gamma^0\gamma^1]\epsilon\ ,
\end{equation}
where the real string-like central charge $Z$ is given by the self dual $H$
charge
\begin{equation}
\int_{\partial{\cal M}}H=2\pi^2Z\ .
\end{equation}
This reinforces the close relation
between the central charges of a supergravity theory and the bosonic charges
of the fields in the graviton multiplet.

{}From the point of view of the supersymmetry algebra, the left hand side of
Eqn.~(\ref{eq:qqanti}) is non-negative for identical (commuting) spinors
$\epsilon'=\epsilon$.  Since $\gamma^0\gamma^1$ has eigenvalues $\pm1$, this
gives rise to the Bogomol'nyi bound
\begin{equation}
GM\ge 2|Z|\ ,
\label{eq:bogo}
\end{equation}
with saturation of the bound corresponding to (partially) unbroken
supersymmetry.  However an issue has arisen over the necessary
conditions for this bound to apply.  In particular, it has been noted
that the gauge dyonic string \cite{gaugestring} may have a tensionless
limit without naked singularities when the instanton size in the gauge
solution is sufficiently large.  Corresponding to
Eqn.~(\ref{eq:bogo}), this tensionless string has vanishing central
charge and is hence quasi-anti-self-dual.  Nevertheless, examination
of the Killing spinor equations indicates that it still breaks exactly
half of the supersymmetries, in contrast to the expectation that $M=0$
yields completely unbroken supersymmetry.  In terms of singular
four-dimensional solutions, this breakdown of the Bogomol'nyi argument
has also been discussed in \cite{Cvetic:1995e,Chan:1995}.

In order to address the issue of where the Bogomol'nyi expression may break
down, we take a closer look at the Witten-Nester proof of the positive
energy theorem \cite{Witten:1981,Nester}.  Following \cite{Dabholkar}, the
charges at infinity may be related to the divergence of Nester's form:
\begin{equation}
\int_{\partial{\cal M}}N^{\mu\nu}d\Sigma_{\mu\nu}=
\int_{\cal M}\nabla_\mu N^{\mu\nu}d\Sigma_\nu\ .
\end{equation}
Proof of the Bogomol'nyi bound is then a matter of reexpressing this
divergence in a manifestly non-negative form.
Straightforward but tedious manipulations allow the divergence of
Nester's form to be rewritten in terms of the supersymmetry variations of
the fermionic fields given in Eqn.~(\ref{eq:susyF}).  Starting with
\begin{eqnarray}
\nabla_\mu N^{\mu\nu}&=&\overline{\delta_{\epsilon'}\psi_\mu}
\gamma^{\mu\nu\rho}\delta_\epsilon\psi_\rho-{\textstyle{1\over2}}
\overline{\epsilon'}G^\nu{}_\sigma\gamma^\sigma\epsilon\nonumber\\
&&+{\textstyle{1\over4}}\overline{\epsilon'}\gamma^{\mu\nu\rho}
\gamma^{\beta\gamma}(\nabla_\mu H_{\rho\beta\gamma})\epsilon
+{\textstyle{1\over16}}\overline{\epsilon'}\gamma^{\beta\gamma}
\gamma^{\mu\nu\rho}\gamma^{\lambda\sigma}H_{\mu\beta\gamma}
H_{\rho\lambda\sigma}\epsilon\ ,
\end{eqnarray}
it is apparent that the $H$ equations of motion must enter the
calculation.  Working through these equations then gives the
final result
\begin{eqnarray}
\nabla_\mu N^{\mu\nu}&=&\overline{\delta_{\epsilon'}\psi_\mu}
\gamma^{\mu\nu\rho}\delta_\epsilon\psi_\rho
+\overline{\delta_{\epsilon'}\chi}\gamma^\nu\delta_\epsilon\chi
+v_rc^r\tr\overline{\delta_{\epsilon'}\lambda}\gamma^\nu
\delta_\epsilon\lambda
-{\textstyle{1\over2}}\overline{\epsilon'}
[G^{\nu\sigma}-{\cal T}^{\nu\sigma}]\gamma_\sigma\epsilon\nonumber\\
&&-{\textstyle{1\over12}}\overline{\epsilon'}\gamma^{\mu\rho}\gamma^\nu
\gamma^{\beta\gamma}[\partial_{[\mu}H_{\rho\beta\gamma]}
+\sqrt{2}P^m_{[\mu}K^m_{\rho\beta\gamma]}-{\textstyle{3\over2}}v_rc^r
\tr F_{[\mu\rho}F_{\beta\gamma]}]\epsilon\nonumber\\
&&+{\textstyle{1\over2}}\overline{\epsilon'}[\nabla_\alpha
H^{\alpha\nu\sigma}-\sqrt{2}P^m_\alpha K^{m\,\alpha\nu\sigma}
-{\textstyle{1\over4}}\epsilon^{\nu\sigma\alpha\beta\gamma\delta}v_rc^r
F_{\alpha\beta}F_{\gamma\delta}]\gamma_\sigma\epsilon\ .
\label{eq:divnester}
\end{eqnarray}
We wish to point out that this is an exact expression, where only
kinematics has been used in rewriting the divergence.  The last two lines
are related to the self-dual $H$ equation of motion (in Bianchi identity
and divergence form respectively), and hence vanish on-shell.  In addition
to the expected terms, this divergence has the unusual feature in that the
full stress tensor ${\cal T}_{\mu\nu}$ arising from the supersymmetry
manipulations is modified by the inclusion of an antisymmetric contribution
\begin{eqnarray}
{\cal T}_{\mu\nu}&=&T_{\mu\nu}+T'_{\mu\nu}\nonumber\\
&=&T_{\mu\nu}-2v_rc^r\tr[F_{\mu\alpha}F_\nu{}^\alpha
-{\textstyle{1\over4}}g_{\mu\nu}F_{\alpha\beta}F^{\alpha\beta}
-{\textstyle{1\over8}}\epsilon_{\mu\nu\alpha\beta\gamma\delta}
F^{\alpha\beta}F^{\gamma\delta}]\ ,
\end{eqnarray}
where $T_{\mu\nu}$, given in (\ref{eq:tmunu}), is the symmetric stress tensor
appearing in Einstein's equation.  In particular, this antisymmetric
component, which arises as a consequence of the $N=(1,0)$ supersymmetry
algebra in six dimensions \cite{Olive}, is related to the fact that
the classical equations of motion, Eqns.~(\ref{eq:eom}), are actually
inconsistent in such a manner as to cancel the effects of the gauge
anomalies when loop corrections are taken into account
\cite{Ferrara:1996,Nishino:1997ff}.
As a result, the equations violate Bose symmetry in a way that would
be impossible if they were derivable from a Lagrangian.  There exists
a Lagrangian, at least in the case $n_T=1$, which automatically leads
to Bose symmetric equations but which lacks gauge invariance
\cite{DMW}.  As discussed in \cite{Ferrara:1996,Nishino:1997ff}, these two
formulations are related to the difference between consistent and
covariant anomalies.  It is interesting to note, however, that the
gauge dyonic string solves both sets of equations, since the Bose
non-symmetric terms vanish in this background.

We are now in a position to examine the conditions under which the
Bogomol'nyi bound, Eqn.~(\ref{eq:bogo}), may hold.  Based on the rewriting
of the Bogomol'nyi equation in terms of a volume integral, it is apparent
that the mass bound will hold provided the divergence
$\nabla_\mu N^{\mu\nu}$ is positive semi-definite over the
entire transverse space ${\cal M}$.
This gives rise to the following three conditions: $i$) the supergravity
equations of motion must be satisfied%
\footnote{Only Einstein's equation and the $H$ equation of motion are
relevant for the Bogomol'nyi calculation.  Note that when we refer to
Einstein's equation, we do not include the correction $T'_{\mu\nu}$ which
is accounted for separately.},
$ii$) Witten's condition must hold globally so the gravitino variation is
non-negative, and $iii$) the Yang-Mills contributions from both the gaugino
variation and the correction $T'_{\mu\nu}$ to the stress tensor must be
non-negative.  While the first condition is straightforward, the other two
require further explanation.  Witten's condition \cite{Witten:1981} is
essentially a spatial Dirac equation, $\gamma^i\delta_\epsilon\psi_i=0$,
where $i=1,\ldots,5$.  While this condition may be satisfied for a well
behaved background, it is also important to ensure that such spinors
are normalizable on all of ${\cal M}$ so that the divergence integral is
well defined.  In particular, this normalizability condition apparently
breaks down in the presence of naked singularities, as we subsequently
verify for the gauge dyonic string solution.  This leads us to believe that
Witten's condition is essentially equivalent to demanding that the
background contains no naked singularities.

We now turn to the conditions that need to be imposed on the Yang-Mills
fields.  Looking at the gaugino variation in Eqn.~(\ref{eq:divnester}), it
is natural to impose the condition that all components of the $n_V$
dimensional vector $v_rc^r$ are to be non-negative.  Since the $n_T$ scalars
encoded in the vielbein $v_r$ act as gauge coupling constants, this condition
simply states that the Yang-Mills fields must have the correct sign kinetic
terms.  Starting from a weakly coupled point in moduli space, it is apparent
that the only way to generate a wrong sign term is to pass through infinite
coupling.  Since this corresponds to a phase transition \cite{DMW}, driven
by tensionless strings \cite{Ganor,Seibergwitten,gaugestring}, it
indicates that the Bogomol'nyi
results need to be applied with care when discussing the strong coupling
dynamics of six dimensional strings.

Since the Yang-Mills fields lead to a modification of the stress tensor, it
is also necessary to require that $T'_{\mu\nu}$ enters non-negatively into
the divergence of Nester's form.  For a string-like geometry in the $0$-$1$
direction, this condition is equivalent to demanding that $-T'_{00} \ge
|T'_{01}|\ge 0$, which is automatically satisfied for gauge fields living
only in transverse space (again provided $v_rc^r$ is non-negative).  To see
this, note that for $\mu,\nu=0,1$ we may write
\begin{equation}
T'_{\mu\nu} = \ft12 v_rc^r\tr[g_{\mu\nu} F_{mn}F^{mn}
+\epsilon_{\mu\nu} F_{mn} *_4 F^{mn}]\ ,
\end{equation}
%
%%% \cite{bpst}
and use the instanton argument, $\tr (F\pm *_4 F)^2\ge0$,
to show that the $T'$ conditions are satisfied.  Therefore
as long as the Yang-Mills fields vanish in the longitudinal directions of
the string-like solution, no further condition is necessary.  It is perhaps
not coincidental that this vanishing of the gauge fields on the string also
renders unimportant the inconsistency of the classical equations of motion.

\subsection{Supersymmetry of the gauge dyonic string}

It is instructive to see how the Bogomol'nyi equation breaks down in the
various limits of the gauge dyonic string.  For this string background,
given by (\ref{eq:gdssol}), the supersymmetry variations of the fermions,
(\ref{eq:susyF}), become
\begin{eqnarray}
\delta\psi_\mu&=&-\gamma^n\partial_nA\gamma_\mu{\cal P}_2^+\epsilon\nonumber\\
\delta\psi_m&=&\gamma^n\partial_nA\gamma_m{\cal P}_2^+\epsilon
+e^{A/2}\partial_m(e^{-A/2}\epsilon)\nonumber\\
\delta\chi&=&-i\gamma^n\partial_n\phi{\cal P}_2^+\epsilon\nonumber\\
\delta\lambda&=&-{1\over2\sqrt{2}}F_{mn}\gamma^{mn}{\cal P}_2^+\epsilon\ ,
\end{eqnarray}
where ${\cal P}_2^+=\half(1+\gamma^{\overline{01}})$ is a projection onto
the chiral two-dimensional world-sheet of the string-like solution
(overlined symbols indicate tangent-space indices).  This
indicates, as noted in \cite{gaugestring}, that the Killing spinor equations
are solved for spinors $\epsilon$ satisfying
\begin{equation}
{\cal P}_2^+\epsilon=0\ ,\qquad\epsilon = e^{A/2}\epsilon_0\ .
\end{equation}
On the other hand, the fermion zero modes are given by spinors $\epsilon$
surviving the projection, namely ${\cal P}_2^+\epsilon=\epsilon$.  Note that
for the zero modes there is no further condition on $\epsilon$.

Based on the above supersymmetry variations, we may explicitly calculate the
divergence of Nester's expression.  Since this expression obviously vanishes
for Killing spinors, we only concern ourselves with the fermion zero modes.
For simplicity in working with the derivative term entering $\delta\psi_m$,
we assume a simple scaling so that $\epsilon$ is given by
\begin{equation}
\epsilon=e^{\alpha A}\epsilon_0\ ,\qquad {\cal P}_2^+\epsilon = \epsilon\ ,
\end{equation}
where $\epsilon_0$ is a constant spinor.  Working out the divergence then
gives
\begin{eqnarray}
\int_{\cal M} \nabla_\mu N^{\mu\nu}d\Sigma_\nu
&=&2\pi^2\, \epsilon_0^\dagger\epsilon_0^{\vphantom{\dagger}}
\int e^{2(\alpha-{1\over2})A}[4(\alpha-\half)\partial_m A
\partial_m A+2\partial_m\partial_m A]r^3 dr\nonumber\\
&=&2\pi^2\, \epsilon_0^\dagger\epsilon_0^{\vphantom{\dagger}}
{1\over\alpha-{1\over2}}
\int r^3dr\partial_m\partial_m e^{2(\alpha-{1\over2})A}\ ,
\end{eqnarray}
where the last line holds for $\alpha\ne\half$ and is in fact a total
derivative, which is not surprising considering the origin of this
expression.  Substituting in the explicit function $A(r)$, we then find
\begin{equation}
2\pi^2\, \epsilon_0^\dagger\epsilon_0^{\vphantom{\dagger}}
[{GM\over2}-Z]=
\int_{\cal M} \nabla_\mu N^{\mu\nu}d\Sigma_\nu
=2\pi^2\, \epsilon_0^\dagger\epsilon_0^{\vphantom{\dagger}}
[Pe^{-\phi_0}+Qe^{\phi_0}]\ ,
\end{equation}
which is independent of $\alpha$ as expected.
Combining this with $[{GM\over 2}+Z]=0$ appropriate to Killing spinors then
gives an explicit derivation of the Bogomol'nyi bound,
\begin{equation}
GM=-2Z=Pe^{-\phi_0}+Qe^{\phi_0}\ ,
\label{eq:gaugebogo}
\end{equation}
for the gauge dyonic string.

So far we have not addressed the issue of what conditions are necessary to
ensure the validity of the Bogomol'nyi bound.  While the equations of motion
are satisfied by construction, both Witten's condition and the positivity of
the gauge function are not guaranteed.  Examining first Witten's condition,
we find
\begin{equation}
\gamma^i\delta_\epsilon\psi_i=\gamma^n\partial_n A
[\alpha-\half-{\cal P}_2^+]\epsilon\ .
\end{equation}
Therefore, for Killing spinors, we choose $\alpha=\half$ as noted previously
in order to satisfy Witten's condition.  On the other hand, we must
choose $\alpha={3\over2}$ for the case of the fermion zero modes.  Provided
there are no naked singularities, this value of $\alpha$ gives rise to a
well-behaved integral, so that there is no problem satisfying Witten's
condition.  However this is no longer the case whenever there are naked
singularities.  To see this, we note that such naked singularities develop
whenever $2Pe^{-\phi_0}\le -\rho^2$ or $2Qe^{\phi_0}\le -\rho^2$ so that
$e^{-2A}$ vanishes for some $r^2\ge 0$ \cite{gaugestring}.  Convergence of
the volume integral near the singularity then requires $\alpha<-{3\over2}$
(or $\alpha<-{1\over2}$ for the case $Pe^{-\phi_0}=Qe^{\phi_0}$) which
clearly indicates the incompatibility of Witten's condition with
normalizable fermion zero modes whenever naked singularities are present.

Note that for any value of the mass given by Eqn.~(\ref{eq:gaugebogo}),
it is always possible to avoid naked singularities in the gauge
dyonic string by choosing a sufficiently large instanton size $\rho$.
Therefore evasion of the Bogomol'nyi bound, Eqn.~(\ref{eq:bogo}), is
possible even without singularities.  Whenever $M<0$ we may see that
the breakdown in Bogomol'nyi occurs because the Yang-Mills couplings have
the wrong sign (this is already obvious because $M$ itself is related to
the gauge coupling at infinity).  A quick check shows that this breakdown
is also present for the tensionless ($M=0$) quasi-anti-self-dual string
where there is an exact cancellation between the contributions from the
graviton and tensor multiplet fields and the wrong sign Yang-Mills fields.
As shown below, this cancellation continues to hold when examining the
energy integral for the tensionless string in the flat-space limit.

\section{The flat-space limit}

If the charges $P$ and $Q$ are such that $P=Q e^{2\phi_0}$, the
anti-self-dual 3-form field strength and the dilaton decouple, {\it i.e.}\
$K^m_{\mu\nu\rho}=0$, $\phi=\phi_0$.  In other words, the matter multiplet
decouples in this case, and we recover the self-dual string of
\cite{Dufflu}.  On the other hand if
$P=-Q e^{2\phi_0}$, the dyonic string becomes massless, as can be seen from
(\ref{massform}).  At first sight, one might think that in this case the
self-dual 3-form $H_{\mu\nu\rho}$ and the metric of the gravity multiplet
would be decoupled.  However, this is not in fact what happens.  This can
easily be seen from the fact that the metric (\ref{eq:gdssol}) does not
become flat: indeed the $1/r^2$ terms cancel asymptotically, as they must
since the solution is now massless, but the metric still has non-vanishing
asymptotic deviations from Minkowski spacetime of order $1/r^4$.  Similarly,
the self-dual 3-form $H_{\mu\nu\rho}$ falls off as $1/r^4$.  On the other
hand, the fields $K^m_{\mu\nu\rho}$ and $\phi-\phi_0$ fall off as $1/r^2$
at large $r$.  For this reason, the dyonic string in this limit should
more appropriately be called quasi-anti-self-dual \cite{gaugestring},
rather than anti-self-dual.  However, the solution becomes anti-self-dual
asymptotically, since the self-dual part of the 3-form falls off faster by
a factor of $1/r^2$.

     The above discussion suggests that it should be possible to take the
flat-space limit of the $N=(1,0)$ supergravity theory, and the
quasi-anti-self-dual solution, where Newton's constant is set to zero.
In fact, as we shall show below, there are actually two distinct limits
that can be taken, yielding two inequivalent flat-space theories.
To show this, we shall first
construct the flat-space limit of the more general $N=1$ supergravity coupled
to an arbitrary number of anti-self-dual fields $K^m_{\mu\nu\rho}$.
To do this, it is convenient to re-introduce Newton's constant
$\kappa$ in the supergravity theory, by rescaling the fields of
the tensor multiplet in the following manner:
\bea
V=\left[\matrix{v_0&  v_M \cr  x^m{}_0& x^m{}_M}\right]
&\longrightarrow &
\left[\matrix{v_0 & \kappa v_M \cr \kappa x^m{}_0& x^m{}_M}\right]\ ,
\label{rescale}\\
B^M_{\mu\nu}& \longrightarrow &\kappa\, B^M_{\mu\nu}\ ,\nonumber\\
\chi^m &\longrightarrow & \kappa\, \chi^m\ ,\nonumber
\eea
while the fields of the Yang-Mills multiplet have not been rescaled, the
coupling constants $c^r$ are naturally dimensionless in the global
limit, and hence must be rescaled according to
\be
c^r \longrightarrow \kappa\, c^r\ .\label{cscale}
\ee
Note that $P^m_\mu\rightarrow \kappa P^m_\mu$ under the rescalings.
As a result of this rescaling, the equations of motion for the
supergravity fields become
\begin{eqnarray}
G_{\mu\nu} -H_{\mu\rho\sigma}H_\nu{}^{\rho\sigma} &=& \kappa^2
[K^m_{\mu\rho\sigma}K_\nu^{m\,\rho\sigma}+2(P_\mu^mP_\nu^m
-\ft12 g_{\mu\nu}P_\rho^mP^{m\,\rho})\nonumber\\
&&+\kappa[4(v_0c^0+\kappa v_Mc^M)\tr
(F_{\mu\lambda}F_\nu{}^\lambda-\ft14 g_{\mu\nu}
F_{\lambda\sigma}F^{\lambda\sigma})]\nonumber\\
dH&=&-\kappa^2\sqrt{2}P^mK^m + \kappa(v_0c^0+\kappa v_Mc^M)\tr F^2\nonumber\\
\gamma^{\mu\nu\rho}\nabla_\nu\psi_\rho+H^{\mu\nu\rho}\gamma_\nu\psi_\rho
&=&\kappa^2[\ft{i}{2} K^{m\,\mu\nu\rho}\gamma_{\nu\rho}\chi^m
-\ft{i}{\sqrt{2}}P_\nu^m\gamma^\mu\gamma^\nu\chi^m]\nn\\
&& -\kappa[\ft{1}{\sqrt{2}}\gamma^{\sigma\tau}\gamma^\mu (v_0c^0+\kappa
v_Mc^M)
\tr F_{\sigma\tau}\lambda]\ ,\label{einst}
\end{eqnarray}
where now $H=v_0{\cal H}^0 + \kappa^2 v_M{\cal H}^M$,
indicating that in the limit $\kappa\to0$ we may consistently set the
gravity fields to their flat-space backgrounds,
\begin{equation}
g_{\mu\nu}\to\eta_{\mu\nu},\qquad
B_{\mu\nu}^0\to0,\qquad
\psi_\mu\to0\ .
\end{equation}
Note that the terms proportional to $c^0$ (the coupling of Yang-Mills
to the self-dual $H$) enter at $O(\kappa)$.  This suggests the
possibility that two different limits can arise; one where $c^0/\kappa$ is
held fixed, and the other where $c^0$ is non-vanishing and held
fixed, as $\kappa$ goes to zero.  This may be made more transparent by
examining the Yang-Mills equation of motion
\begin{eqnarray}
D^\mu[(v_0c^0+\kappa v_Mc^M) F_{\mu\nu}]&=&
H_{\nu\rho\sigma}(v_0c^0+\kappa v_Mc^M)F^{\rho\sigma}
+ \kappa K_{\nu\rho\sigma}^m(\kappa x^m{}_0c^0+x^m{}_Mc^M)F^{\rho\sigma}
\ ,\nonumber\\
\label{ymeom}
\end{eqnarray}
from which we see that the $O(\kappa^0)$ terms survive only in the second
limit, whilst the equation is of order $\kappa$ in the first limit.
Before proceeding with the flat-space limits, we note that the
constrained vielbein matrix $V$ simplifies greatly in the $\kappa\to0$
limit, and the $n_T$ degrees of freedom can be parametrized by scalar
fields $\phi^m$ defined by $\delta^m_Mv_M=x^m{}_0=\phi^m$.  The other
components of $V$ simply become $v_0=1$ and $x^m{}_M =\delta^m_M$.
The two flat-space limits arise as follows:

\bigskip
\noindent {\underline{\it Flat-space limit with $c^0/\kappa$ fixed:}}
\medskip

In this limit, it is natural to define $\tilde c^0=c^0/\kappa$ before
taking the flat-space limit.   We see that there are now no
$\kappa$-independent terms in (\ref{ymeom}), and we obtain a
Yang-Mills equation that includes interactions
with the anti-self-dual matter multiplets.  We find that the complete set of
flat-space equations is
\bea
\Box \phi^m &=& c^m\, \tr(F_{\mu\nu}F^{\mu\nu})\ ,\nonumber\\
\del^\mu \, K^m_{\mu\nu\rho} &=& -\ft14 c^m \,
\epsilon_{\mu\nu\rho\a\b\gamma}\, \tr(F^{\mu\a} F^{\b\gamma})\ ,\nonumber\\
\gamma^\mu\del_\mu \chi^m &=& -\ft{i}{\sqrt2} c^m\, \tr( F_{\mu\nu}
\gamma^{\mu\nu} \lambda)\ ,\nonumber\\
D^\mu[ (\tilde c^0 + c^m\, \phi^m) \, F_{\mu\nu}] &=& c^m \, F^{\rho\sigma}\,
K^m_{\nu\rho\sigma} \ ,\label{flateqs}\\
(\tilde c^0+c^m \phi^m )\gamma^\mu D_\mu \lambda
&=& -\ft12 c^m (\del_\mu \phi^m)
\gamma^\mu \lambda -\ft{i}{2\sqrt2} c^m\, F_{\mu\nu} \gamma^{\mu\nu} \chi^m
- \ft1{12}c^mK^m_{\mu\nu\rho}\gamma^{\mu\nu\rho}\lambda\ .\nonumber
\eea
Note that the anti-self-dual field strengths are given by
\be
K^m = dB^m + c^m\, \omega_3\ ,\label{kcs}
\ee
where $\omega_3=AdA + \ft23 A^3$. The supersymmetry transformation
rules in this flat-space limit become
\bea
&&\delta \phi^m =\bar\epsilon \chi^m\ ,\qquad \delta A_\mu=-
\ft{i}{\sqrt2}\bar\epsilon\gamma_\mu\lambda\ ,
\qquad \delta B^m_{\mu\nu}=\ft12 \bar\epsilon \gamma_{\mu\nu} \chi^m +2
c^m \tr(A_{[\mu} \delta A_{\nu]})\ ,\nonumber\\
&&\delta\lambda = -\ft{1}{2\sqrt2}\, F_{\mu\nu} \gamma^{\mu\nu} \epsilon
\ , \qquad \delta \chi^m = -\ft{i}{2} \del_\mu \phi^m\, \gamma^\mu \epsilon
+\ft{i}{12} K^m_{\mu\nu\rho}\, \gamma^{\mu\nu\rho} \epsilon\ .
\label{flatsusy}
\eea

      The energy-momentum tensor for this flat-space theory may be
obtained simply by applying the same limiting procedure to the
right-hand side of the Einstein equation of the original supergravity
theory, given in (\ref{einst}).  By this means we obtain the
flat-space expression
\be
T_{\mu\nu}=K^m_{\mu\rho\sigma}\, K^m_\nu{}^{\rho\sigma} +
\del_\mu\phi^m\,
\del_\nu\phi^m -\ft12 \eta_{\mu\nu}\,(\del\phi^m)^2 +
4(\tilde c^0 + c^m\, \phi^m)\tr(F_{\mu\lambda}\, F_\nu{}^\lambda
-\ft14 \eta_{\mu\nu} \, F_{\lambda\sigma}\, F^{\lambda\sigma})
\ .\label{stress1}
\ee

This is the theory that we refer to as the ``interacting theory".
It is interesting to note that the bosonic equations of motion of
(\ref{flateqs}) can be derived from the Lagrangian
\be
{\cal L} = -(\del\phi^m)^2 -\ft13 K^2 -2(\tilde c^0 + c^m\, \phi^m)\,
\tr(F^2) -2c^m\, {*(B^m\wedge \tr(F\wedge F))}\ ,\label{grandma}
\ee
where $K$ is taken to be unconstrained, with its anti-self-duality
being imposed only after having obtained the equations of motion.

\bigskip
\noindent {\underline{\it Flat-space limit with $c^0$ held fixed:}}
\medskip

      The situation is different when $c^0$ is non-vanishing and is
held fixed when $\kappa$ goes to zero.  As
can be seen from (\ref{ymeom}), the leading-order terms in the
Yang-Mills equation are now independent of $\kappa$, and in fact there
are now no interactions with the anti-self-dual multiplets in the
$\kappa\to 0$ limit.  All equations of motion, and supersymmetry
transformation rules, remain the same as in the previous $c^0\sim
\kappa$ limit with the exception of the Yang-Mills equations and the
gaugino equation, which are now source-free and given by
\bea
D^\mu\, F_{\mu\nu} =0\ ,\nn\\
\gamma^\mu\, D_\mu\, \lambda = 0\ .\label{free}
\label{eq:bsseom}
\eea
Note however that the energy-momentum tensor, again obtained from the
right-hand side of the Einstein equation in (\ref{einst}) by applying
the limiting procedure, is now simply given by
\be
T_{\mu\nu}= 4c^0\, \tr(F_{\mu\lambda}\, F_\nu{}^\lambda
-\ft14 \eta_{\mu\nu} \, F_{\lambda\sigma}\, F^{\lambda\sigma})
\ .\label{stress2}
\ee
In the case of a single self-dual tensor multiplet, this is the theory
that we refer to as the ``BSS theory''.

A word of explanation is in order here.  Firstly, it should be noted
that this energy-momentum tensor arose as a term of order $\kappa$ in
the Einstein equation, rather than the usual order $\kappa^2$ for
matter fields.  Consequently the Yang-Mills contribution dominates the
$O(\kappa^2)$ contributions from the tensor multiplets, and so they
are absent in this flat-space limit.  Indeed, it is evident that if
one were to add ``standard'' contributions for the fields of the
tensor multiplets, one would find that the resulting energy-momentum
tensor was not conserved upon using the equations of motion.
Effectively the tensor multiplets describe ``test fields'' in a
Yang-Mills background, whose energy-momentum tensor is negligible in
comparison to that of the Yang-Mills field.  For the same reason, they
do not affect the Yang-Mills equation.  The energy-momentum tensor
(\ref{stress2}) would cease to be appropriate in a configuration where
the Yang-Mills field was zero, since now the previously-neglected
matter contributions would become important.  This rather pathological
feature of the BSS theory is reflected also in the fact that it cannot
be described by an analogue of the Lagrangian (\ref{grandma}), owing to
the inherent asymmetry between the occurrence of interaction terms in
the matter and Yang-Mills equations.

    A number of further comments are also in order.  Firstly, it should
be emphasised that the higher-order fermi terms are not included in
the equations of motion and supersymmetry transformation rules
(\ref{flateqs}) and (\ref{flatsusy}); they were not included in
\cite{RomansSD,Sagnotti}, and indeed they have only recently been
computed \cite{Riccioni}. (See also
\cite{Nishino:1997ff,Tonin}.)
Nevertheless, one can see on general grounds that the inclusion of the
higher-order terms in the supergravity theory will not present any
obstacle in the taking of the two inequivalent flat-space limits.  
Alternatively, the higher-order completion of the supersymmetry
transformations may be determined in either of the flat-space theories
by demanding the closure of the supersymmetry algebra on the fermi
fields.  For the interacting theory the supersymmetry transformation rules
for the bosons remain unchanged, while the complete transformation rules
for the fermions are
\begin{eqnarray}
\delta\chi^m &=& -\ft{i}{2}[\partial_\mu\phi^m\gamma^\mu - \ft{1}{6}
K_{\mu\nu\rho}^m\gamma^{\mu\nu\rho}]\epsilon
-\ft12c^m\tr [\gamma_\mu\lambda(\overline{\epsilon}\gamma^\mu\lambda)]
\label{eq:hosusy}
\ ,\\
\delta\lambda &=& -\ft{1}{2\sqrt{2}}F_{\mu\nu}\gamma^{\mu\nu}\epsilon
+{c^m\over (\tilde
c^0+c^n\phi^n)}\big[-\ft12(\overline{\chi}^m\lambda)\epsilon
-\ft14(\overline{\chi}^m\epsilon)\lambda + \ft18
(\overline{\chi}^m\gamma_{\mu\nu}\epsilon)\gamma^{\mu\nu}\lambda\big]
\ ,\nonumber
\end{eqnarray}
and agree with the flat-space limit of the transformations in the
supergravity theory \cite{Riccioni}.
On the other hand, in the BSS theory the gaugino variation remains
unmodified, and only $\delta\chi^m$ picks up a higher-order correction
(identical to that of the interacting theory).  Note that it is
straightforward to see that this must be the case, since the lowest-order
transformation for the Yang-Mills multiplet, (\ref{flatsusy}), already
closes on the source-free gaugino equation of motion, (\ref{eq:bsseom}).
So in fact we see that the only difference in the supersymmetry
transformation rules in the two flat-space limits is in the higher-order
terms in the gaugino variation, consistent with the difference in the
equations of motion for the Yang-Mills multiplet between the two limits.

The complete gaugino transformation rule in the interacting theory is
somewhat unusual, in that it contains a possibly singular denominator,
$(\tilde c^0 + c^n\phi^n)$ (which was also noted in
\cite{Nishino:1997ff,Riccioni}).  As in the supergravity situation, this
singular denominator is just a manifestation of the strong coupling
singularity already present in the lowest-order Yang-Mills equations.
This form of the denominator also shows up in the complete equations
of motion, given for the fermi fields in the interacting theory by%
\footnote{While these equations of motion were obtained by taking the
flat-space limit of \cite{Riccioni}, they equally well follow from closure
of the supersymmetry algebra, (\ref{eq:hosusy}).}
\begin{eqnarray}
\gamma^\mu\del_\mu \chi^m &=& -\ft{i}{\sqrt2} c^m\, \tr( F_{\mu\nu}
\gamma^{\mu\nu} \lambda)
+ {i c^mc^n\over (\tilde c^0 + c^p\phi^p)}
\tr \big[\ft32(\overline{\chi}^n\lambda)\lambda -\ft14
(\overline{\chi}^n\gamma_{\mu\nu}\lambda)\gamma^{\mu\nu}\lambda\big]
\ ,\nonumber\\
(\tilde c^0+c^m \phi^m )\gamma^\mu D_\mu \lambda
&=& -\ft12 c^m (\del_\mu \phi^m)
\gamma^\mu \lambda -\ft{i}{2\sqrt2} c^m\, F_{\mu\nu} \gamma^{\mu\nu} \chi^m
- \ft1{12}c^mK^m_{\mu\nu\rho}\gamma^{\mu\nu\rho}\lambda\nonumber\\
&& -i {c^mc^n\over (\tilde c^0 + c^p\phi^p)}
\big[\ft34(\overline{\lambda}\chi^m)\chi^n-\ft18
(\overline{\lambda}\gamma_{\mu\nu}\chi^m)\gamma^{\mu\nu}\chi^n\big]
\nonumber\\
&&+i\alpha c^mc^{m\prime}\tr'[(\overline{\lambda}\gamma_\mu\lambda')
\gamma^\mu \lambda'] \ ,
\end{eqnarray}
where the primes in the last line indicate the quantities involved in
the trace.  (Recall that there can be different $c^m$ constants for
each factor in a semi-simple group.)  Note that $\alpha$ is an
arbitrary parameter that is not fixed by the supersymmetry algebra
\cite{Riccioni}, and appears to be related to the gauge anomaly (see
\cite{Riccioni} for a more complete discussion).

We also 
note that the flat-space limit when $c^0$ is non-vanishing and held
fixed, if we specialise to the case where there is only a single
anti-self-dual multiplet, coincides with the BSS theory, constructed in
\cite{Bergshoeff:1996qm}.  It was argued in \cite{Bergshoeff:1996qm}
that this theory could not be obtained as a $\kappa\to 0$ limit of the
supergravity theory, on the grounds that the Chern-Simons form $\omega$
enters the 3-form field strengths in (\ref{csterms}) with a factor of
$\kappa$ (after restoring Newton's constant, as in (\ref{cscale})), and
thus it would disappear in the flat-space limit.  However, while this
is indeed the case for the self-dual field of the gravity multiplet,
the potentials
$B^m_{\mu\nu}$ for the anti-self-dual matter fields also acquire
factors of $\kappa$, with the net result that the Chern-Simons terms
are of the same order, and hence they survive in the $\kappa\to 0$
limit, as we saw in (\ref{kcs}) above.  The non-standard dimensions of
the energy-momentum tensor (\ref{stress2}) is a reflection of the need
for a dimensionful free parameter, which was also seen in
\cite{Bergshoeff:1996qm}.  Finally, we remark that the more general
flat-space theory we obtained in the limit where $c^0 \sim \kappa$,
does not conflict with the results in \cite{Bergshoeff:1996qm} which
found only the free Yang-Mills equations (\ref{free}), since in
\cite{Bergshoeff:1996qm} it was assumed that the kinetic term for the
Yang-Mills multiplet was described by the standard superspace free
action.  Note also that the BSS theory can be obtained from the
interacting flat-space theory by taking $k$ to zero after making
the following rescalings of the fields of the interacting theory:
\be
\phi^m\to k\, \phi^m\ ,\qquad B^m_{\mu\nu} \to k\, B^m_{\mu\nu}\ ,
\qquad \chi^m\to k\, \chi^m\ ,
\ee
together with the rescaling $c^m\to k\, c^m$.  Thus the interacting
flat-space theory encompasses the BSS theory as a singular limiting
case.

     Let us now consider the flat-space limit of the
quasi-anti-self-dual dyonic string solution (\ref{eq:gdssol}) of the
supergravity theory.   This solution is massless, and hence from
(\ref{massform}) it follows that the magnetic charge is related to the
electric charge by $P=-Q\, e^{2\phi_0}$.  Consequently, the parameters
$c^0=(Q+P)/32$ and $c^1=(Q-P)/32$ are given by
\be
c^0= - \ft1{16} Q\, e^{\phi_0}\, \sinh \phi_0 \ ,\qquad
c^1 = \ft1{16} Q\, e^{\phi_0}\, \cosh \phi_0\ .
\ee
In the flat-space limit, where in particular $\phi$ was rescaled by
$\kappa$, we see that $c^0 =-\ft1{16} Q\, \kappa\, \phi_0$ prior to
sending $\kappa$ to zero, and hence we are in the regime of the
``interacting theory'', corresponding to the first of the two limits
discussed above.  We find that the flat-space solution is given by
\begin{eqnarray}
&&\phi=\phi_0 - \fft{Q(2\rho^2+ r^2)}{(\rho^2 +r^2)^2} \nonumber\\
&&K_{mnp} = -\half\epsilon_{mnpq}\partial_q\phi\ ,\qquad
K_{\mu\nu m} = -\half\epsilon_{\mu\nu}\partial_m\phi\ ,\nonumber\\
&&F^a = \fft{2\rho^2}{(\rho^2+r^2)^2}\, \eta^a_{mn}\,
dy^m\wedge dy^n\ .
\label{eq:flatgds}
\end{eqnarray}
Note that $\phi_0$ no longer has physical significance as a coupling
constant, and it can be eliminated by making a constant shift of $\phi$.

    Since this solution has been obtained as the flat-space limit of a
tensionless string, we expect that it should have vanishing energy.
This might at first sight seem surprising, since it is described by a
non-trivial field configuration.  However, a straightforward
calculation of $T_{00}$ given by (\ref{stress1}) yields
\bea
T_{00}&=& K^2_{00} +\ft12 (\del\phi)^2 + \ft1{16} Q\, (\phi-\phi_0)\,
\tr(F^2)\ ,\nonumber\\
&=& \fft{4 Q^2\, r^2\, (3\rho^2+r^2 )^2}{(\rho^2+ r^2)^6} -\fft{24
Q^2\, \rho^4\, (2\rho^2+r^2)}{(\rho^2+r^2)^6}\ ,\label{t00}
\eea
where the first term in the second line comes from the (equal)
contributions from $K$ and $\phi$, and the second term comes from
$F$.  It is easily verified that while $T_{00}$ itself is
non-vanishing, the integral $\int_0^\infty  T_{00}\, r^3\, dr$ is
equal to zero.  Clearly the Yang-Mills field is giving a negative
contribution to the energy, in precisely such a way that the total
energy is zero.  This is the flat-space analogue of the cancellation
that occurs in the supergravity theory, with its associated subtleties
in the Bogomol'nyi analysis, which we discussed at the end of section~3.

     It should be emphasised that the vanishing energy of the
flat-space tensionless string occurs for arbitrary scale size $\rho$
of the Yang-Mills instanton.  However, if we consider instead the
neutral tensionless string, which can be achieved by setting $\rho=0$
so that the instanton is not present, then the expression (\ref{t00})
becomes $T_{00}=4Q^2/r^6$, whose integral over the transverse space
diverges at the core of the string.  Thus the Yang-Mills instanton in
the gauge-dyonic string can be viewed as a regulator for the total
energy.

     There are also massive string solutions to the interacting
flat-space theory, which can also be obtained as flat-space limits of
the curved-space gauge dyonic string. They arise by taking the ADM
mass, as given by (\ref{massform}), to be non-zero and of the form
$m_0\, \kappa$.  Upon taking the $\kappa\to0$ flat-space limit, this
gives a solution of the same form as (\ref{eq:flatgds}), but with
$\phi$ shifted by the constant $m_0$.  From (\ref{t00}), this gives an
extra term in $T_{00}$ which gives rise to an energy $m_0$ per unit
length for the flat-space string.

    It is interesting to note that while the flat-space limit of the
tensionless string always results in the $c^0\sim\kappa$ limit of the
interacting theory, the final solution itself, as given in
(\ref{eq:flatgds}), also satisfies the equations of motion of the BSS
theory, where $c^0$ is held fixed in the flat-space limit%
\footnote{The solution (\ref{eq:flatgds}) has also been obtained in
the BSS theory by directly solving its first-order BPS equations
\cite{nairprivate}.}.
To see this, we note that for a bosonic background, only the Yang-Mills
equation differs between the two flat-space theories.  In particular,
both Yang-Mills equations may be expressed as
$D^\mu F_{\mu\nu}=J_\nu$, where the current is
\begin{equation}
J_\mu=[c^mF_{\mu\nu}\partial^\nu\phi^m
+c^mF^{\rho\sigma}K^m_{\mu\rho\sigma}]
/[\tilde c^0+c^n\phi^n]\ ,
\label{eq:current}
\end{equation}
for the first theory, and vanishes for the latter.  Because of the
form of the solution, (\ref{eq:flatgds}), we see that $J_\mu$
identically vanishes, and hence the background is indeed a solution to
both flat-space limits.  Furthermore, examination of the BPS
conditions arising from (\ref{flatsusy}) indicates that $J_\mu=0$
for any string-like background preserving half of the supersymmetries.
It should be remarked, however, that when interpreted as a solution to
the $c^0$ fixed flat-space limit, the string no longer has vanishing
energy per unit length, since in this case the stress tensor
(\ref{stress2}) has only a positive contribution from the Yang-Mills
instanton.  Thus only the interacting theory from the first flat-space
limit, (\ref{flateqs}), provides a suitable description of the
tensionless string in flat space.

Finally, we note that by taking the divergence of (\ref{eq:current}),
we obtain
\begin{equation}
D^\mu J_\mu = \ft18 c^mc^{m\prime}
\epsilon_{\mu\nu\rho\sigma\eta\lambda}F^{\mu\nu}
\tr' F^{\rho\sigma\prime}F^{\eta\lambda\prime}/
[\tilde c^0+c^n\phi^n]\ ,
\end{equation}
indicating that the current is not conserved classically.  Thus the
inconsistency of the supergravity theory, which we discussed in
section 3,
survives in the ``interacting'' flat-space limit.  Nevertheless since,
as for the gauge dyonic string in curved space,
$J_\mu$ vanishes identically for the global gauge string, this classical
inconsistency does not spoil the solution.  On the other hand, since the BSS
theory is free of this inconsistency it is possible that such a
classical inconsistency, necessary for anomaly cancellation in the
quantum theory, is an integral part of a fully interacting theory.

We have not paid much attention in this paper to the question of 
gravitational anomalies 
which is always an important issue when dealing with {\it chiral} 
theories. In particular, we have for simplicity ignored the presence 
of hypermultiplets. A coupled supergravity-matter theory which is initailly 
free of gravitational anomalies theory will not remain so when the 
gravity multiplet is switched off because the contributions from the 
gravitino and self-dual $2$-form, necessary for the anomaly cancellation, 
are no longer present. Naively, of course, one could argue that 
gravitational anomalies are no longer of any concern in the flat space 
limit. However, it may be that subtleties arise when one tries to take 
the global limit of a worldvolume theory which relies for its anomaly 
freedom on anomaly inflow from the bulk. This is deserving of further 
study.

    In conclusion, we note that six-dimensional global models are also
important as fivebrane worldvolume theories
\cite{Dijkgraaf,Seiberg,Schwarz,Bergshoeff:1996qm}.  In the absence of
Yang-Mills fields, the $(1,0)$ multiplet is the only one available to
describe the worldvolume theory of the $D=7$, $N=1$ fivebrane solution
found in \cite{lps}.  Six-dimensional global models also arise from 
configurations of higher-dimensional branes with six worldvolume 
dimensions in common. Indeed, the brane configurations yielding $(1,0)$ 
theories with tensor multiplets, 
vector multiplets and hypermultiplets have been identified in 
\cite{Hanany,Intriligator}, although no field equations were written 
down. Here we speculate further that the interacting
anti-self-dual-tensor Yang-Mills system given in this paper (together 
with hypermultiplets where necessary) is the  appropriate one to describe 
these global models.
The global gauge anti-self-dual string,
and in particular the tensionless string, could then be regarded as
strings on the worldvolume. 

\bigskip
\noindent{\bf Note added}

 Global $D=6$, $(1,0)$ models of the type discussed in this paper have 
 recently been shown to arise from configurations of NS fivebranes, 
 Dirichlet sixbranes and eightbranes \cite{Hanany2}

\bigskip
\noindent {\bf Acknowledgment.}

MJD thanks E. Sezgin, E. Verlinde and E. Witten, and JTL  thanks V.P. Nair, for
useful discussions.

\vfill\eject
%%%%%%%%%%%%% include bibtex generated bibliography %%%%%%%%%%%%%%%

%%%%%%%%%%%%%%%%%% end bibliography %%%%%%%%%%%%%%%%%%%%%%%%%%%%%%%

\end{document}